\documentclass[aps,twocolumn,epsfig,eqsecnum]{revtex4}
\newcommand\be{\begin{eqnarray}}
\newcommand\ee{\end{eqnarray}}
\newcommand\ba{\begin{array}}
\newcommand\ea{\end{array}}

\def\cH{{\cal H}}
\def\cS{{\cal S}}

\def\cI{{\cal I}}

\def\cE{{\cal E}}

\begin{document}
\title{Notes on optimality of direct characterisation of quantum dynamics}
\author{M\'ario Ziman$^{1,2}$}
\address{
$^{1}$Research Center for Quantum Information, Slovak Academy of Sciences,
D\'ubravsk\'a cesta 9, 845 11 Bratislava, Slovakia \\
$^{2}$ {\em Quniverse}, L{\'\i}\v{s}\v{c}ie \'{u}dolie 116, 841 04
Bratislava, Slovakia\\
}
\begin{abstract}
We argue that the claimed optimality of a 
new process tomography method suggested in \cite{lidar} is 
based on not completely fair comparison that does 
not take into account the available information in an equal way. 
We also argue that the method is not a new process 
tomography scheme, but rather represents an interesting modification
of ancilla assisted process tomography method. In our opinion these
modifications require deeper understanding and further investigation.
\end{abstract}
\maketitle


\section{Problem}
Authors in \cite{lidar} 
proposed a scheme for process reconstruction, which in their
opinion is more optimal than the other schemes. Here we would like
to question this optimality and even several other points made by 
authors. The following classification 
of different reconstruction schemes is used (according to \cite{lidar}):
\begin{enumerate}
\item{SQPT (Standard quantum process tomography)}
In this scheme one analyze how a 
collection of test states (i.e. a collection
of linearly independent states $\varrho_k\in\cS(\cH_S)$ forming a basis 
of the set of operators) is affected by the action of the 
unknown device. In particular, as a result
one gets a set of assignements $\varrho_k\to\varrho_k^\prime$ that
due to linearity of quantum dynamics specify the channel completely. 
\item{AAPT (Ancilla-assisted process tomography)} 
In this scheme we employ a suitable ancilliary system. In particular,
we prepare a state $\Omega\in\cS(\cH_S\otimes\cH_A)$, we let the device acts 
on the system part of the state (ancilla evolves either trivially, or in some 
known way), and finally, we perform the measurements on the output joint
state $\Omega^\prime=\cE\otimes\cI[\Omega]$. With a suitable choice of 
the input state (it must be from the family of so called
faithful states \cite{dariano})
one can deduce the action of the device $\cE$ 
from the single state assignement $\Omega\to\Omega^\prime$.

\item{DCQD (Direct characterization of quantum dynamics)} 
This scheme is proposed by authors. It is based
on AAPT scheme and a simple idea that instead of doing 
several measurements of the output state one can use a fixed single
measurement, but preparing several different input states. This enables us
to access to all the parameters of the quantum device in a similar
way as in the previous schemes, i.e. we are able to deduce tha action
of $\cE$ on operators forming the operator basis.
\end{enumerate}

\subsection{First comment: Quantification of resources.} 
Authors conclude that their scheme (DCQD) is more optimal than
the remaining two. This result is illustrated in the Table II of their 
paper. They claim the following (for $n$ qubits):\\
\begin{center}
\begin{tabular}{|c|c|c|c|l}
\hline
scheme & $N_{\rm in}$ & $N_{\rm out}$ & $N_{\rm total}$ \\  
\hline
SQPT & $4^n$ & $4^n$ & $16^n$ \\
\hline
AAPT & 1 & $16^n$ & $16^n$ \\
\hline
DCQD & $4^n$ & 1 & $4^n$ \\
\hline 
\end{tabular}
\end{center}

In this table $N_{\rm in}$ stands for number of  
ensembles of different input states, $N_{\rm out}$ stands for  number of
output measurements per one ensemble, and $N_{\rm total}=N_{\rm in}
N_{\rm out}$ is the total number of measurements in the scheme. 
The optimality of DCQD is based on the comparison of $N_{\rm total}$
for all schemes. However, this table is not only not completely precise, 
and also not completely fair. 
Each quantum device acting on $d$ dimensional quantum system is
described by $d^2(d^2-1)$ independent parameters that has to be
specified in arbitrary (complete) process tomography scheme. For 
$n$ qubits this means that we need to specify $N=2^{2n}(2^{2n}-1)=16^n-4^n$.
This is exactly the number of independent parameters we are specifying in all
the mentioned schemes. 

Let us get back to the main point of our comment that
the authors did not correctly quantify the minimal number of 
measurements needed for particular tomography schemes. For
SQPT they calculated the number of single qubit 
measurements. Each qubit measurement 
effectively provides us with only one number (mean value)
that corresponds to one the parameters of the device map $\cE$.
The same logic is used also for calculating the number
of measurements in the case of AAPT. Again they assume that
each measurement (called by authors joint single qubit measurement)
specify only one number. But this is not true, because
each measurement now has $2^n$ different results and the measured
probability distribution contains $2^n-1$ independent parameters.
As we will see exactly this effect stands behind the claimed
optimality of DCQD method, because in that case they consider
Bell-state measurements and take into account the information contained
in the whole probability distribution (not only in mean value). 
Therefore, this measurement allows authors to gain more 
information about the transformation, which finally results in the smaller 
total number of measurements (in comparison with AAPT scheme not treated 
in the same way). From this point of view
the conclusion is not that surprising, since they did not 
use the available resources in the same manner. The problem 
is that for SQPT scheme for $d$ dimensional
system each measurement gives $d-1$ independent numbers, whereas
for AAPT and DCQD each measurement gives $d^2-1$ independent 
numbers. For suitable choices of measurements these numbers
specifies the corresponding number of parameters 
of the device map $\cE$.
The correct version of the table for a single qubit 
(the generalization for arbitrary dimensional system is obvious) 
is the following:\\
\begin{center}
\begin{tabular}{|c|c|c|c|l}
\hline
scheme & $N_{\rm in}$ & $N_{\rm out}$ & $N_{\rm total}$ \\  
\hline
SQPT & $4$ & $3$ & $12$ \\
\hline
AAPT & 1 & $4$ & $4$ \\
\hline
DCQD & $4$ & 1 & $4$ \\
\hline 
\end{tabular}
\end{center}

Thus, we see that AAPT and DCQD are (after proper evaluation)
of the same quality. However, also
such conclusion is, in our opinion, not correct and fair.
To really compare the resources for different tomography schemes
one would need to be able to quantify the complexity of
the preparations and measurements. In our opinion such
quantification will result in mutual equivalence of all
these schemes. To see some differences one should indeed include
and discuss particular physical systems, or discuss 
process reconstructions from incomplete data, or small statistical 
samples.

In the above table it is easy to see that each scheme
provides us with the same number of numbers. We remind that
for AAPT and DCQD schemes the measurements have 4 outcomes, i.e.
the measured probabilities contain 3 independent numbers. 
Therefore, for total number of scpecified parameters
we obtain: for AAPT 1x4x3=12 and for DCQD 4x1x3=12. This is not
surprising and only means that there is no redundant
information in all of the schemes.

Finally, let us consider POVM instead of von Neumann measurements
and perform similar comparison of reseources.
In this case the state tomography can be accomplished by 
a single so called informationally complete POVM, i.e. $N_{\rm out}=1$
for all possible reconstruction schemes. The quantification of
the quality of schemes in the number of POVMs would lead to
the following table for the qubit:\\
\begin{center}
\begin{tabular}{|c|c|c|c|l}
\hline
scheme & $N_{\rm in}$ & $N_{\rm out}$ & $N_{\rm total}$ \\  
\hline
SQPT & $4$ & $1$ & $4$ \\
\hline
AAPT & 1 & $1$ & $1$ \\
\hline
DCQD & $4$ & 1 & $4$ \\
\hline 
\end{tabular}
\end{center}

Thus the DCQD is as efficient as SQPT scheme and AAPT scheme is
much better when one calculate the resources in terms of realized
POVMs. However, let us remind that we do not think that this
is the way how to compare different reconstruction schemes.
In fact, to perform the reconstruction experimentally
one really needs to perform the tomography of input states
as well. This also should be counted among the needed
resources. 

\subsection{Second comment: usage of quantumness.}
In our view the DCQD scheme is just a modification of
the AAPT scheme, in which instead of performing several 
measurements we use different preparations. We do not share
the opinion of authors that only in DCQD scheme
the quantumness is used. The meaning of quantumness
is equivalent to to the necessity of entangled input states 
and entangled measurements in the scheme.
Similarly, like AAPT scheme can be realized with, 
or without entanglement, the same holds also for DCQD method. 

\subsection{Third comment: direct state tomography.}
Authors claimed that the suggested DCQD scheme access
the process directly, i.e. without any usage of state 
tomography. This is indeed true that in this scheme we do not
have sufficient knowledge about the output state, and therefore
we can perform only partial state tomography.
In fact, neither for AAPT scheme we really perform complete
state tomography, since we need to specify only
$d^4-d^2$ parameters instead of $d^4-1$ required for $d^2$
dimensional system. The others parameters describe
the local state of the ancilla system which is not changed 
and known. We agree that in AAPT and SQPT one can in principle perform 
complete state tomography, but we do not think this is somehow important.
To learn anyhting about the quantum process one needs to performs
measurements on states. It is not surprising that one can 
do even more, i.e. we can exclude also the complete knowledge 
about the input state preparators. It is sufficient only to analyze 
(for a fixed measurement) the transformation of 
input/output probability distributions. However, we would
not call it new process tomography method. In our opinion
there are logically only two different methods: 
1) SQPT and 2) AAPT. It is our choice how we will
use the particular resources, i.e. the preparators and 
the measurements. A similar process tomography method
(as DCQD scheme is related to AAPT scheme) can be designed 
also for SQTP scheme. That is, the fact that we have access to
particular state assignements (state tomography) 
is not that important in the process tomography method
and does not represent some redundant information.

\section{Conclusion}
Let us conclude that in our opinion the authors did not 
succeed to show that their scheme is more optimal. It seems that the scheme
can be more efficient in cases when we do not have the opportunity to
perform all measurements. However, since the tomography of input states
(callibration of preparators) is a necessary part of process 
tomoraphy, these measurements (allowing complete state tomography)
must be available. It can be very difficult to
experimentally perform some measurements and only in this sense the
DCQD scheme can be more optimal. The modifications of SQPT and AAPT
schemes are very interesting options that can be very useful in
particular process estimation problems (for instance incomplete
reconstructions) and therefore they deserve further investigation. 
The proposed DCQD scheme represents an interesting example within 
such program and provides an important estimation algorithm that
can be very useful and optimal in particular physical realizations.

\acknowledgements 
I wish to thank Vladim\'\i r Bu\v zek for discussion and
Jason Twamley for his hospitality during my visit in Macquarie 
university. This work was supported in part by the European
Union  projects QAP and by the project APVT-99-012304.

\end{document}